\documentstyle[lutypaper,mydefs,12pt]{article}

\newcommand{\fund}{\drawsquare{6.5}{0.4}}

\newcommand{\asymm}{\raisebox{-3pt}{\drawsquare{6.5}{0.4}\hskip-6.9pt%
        \raisebox{6.5pt}{\drawsquare{6.5}{0.4}}}}

\newcommand{\asymmfour}{\raisebox{-10pt}{\drawsquare{6.5}{0.4}}\hskip-6.9pt%
\raisebox{-3.5pt}{\drawsquare{6.5}{0.4}}\hskip-6.9pt%
\raisebox{3pt}{\drawsquare{6.5}{0.4}}\hskip-6.9pt%
        \raisebox{9.5pt}{\drawsquare{6.5}{0.4}}}

\newcommand{\bb}[1]{\hbox{\bf #1}}

\newcommand{\vu}{\avg{\bar{U}}}

\begin{document}

\begin{titlepage}

\preprint{SLAC-PUB-7714\\
UMDHEP 98-68\\
UCB-PTH-97/62\\
LBNL-41148}

\title{Composite Quarks and Leptons from\\\medskip
Dynamical Supersymmetry Breaking\\\medskip
without Messengers}

\author{Nima Arkani-Hamed}

\address{Stanford Linear Accelerator Center,
Stanford University\\
Stanford, California 94309, USA\\
{\tt nima@slac.stanford.edu}}

\author{Markus A. Luty\footnote{Sloan Fellow}}

\address{Department of Physics,
University of Maryland\\
College Park, Maryland 20742, USA\\
{\tt mluty@physics.umd.edu}}

\author{John Terning}

\address{Department of Physics,
University of California\\
Berkeley, California 94720, USA\\
and\\
Theory Group, Lawrence Berkeley Laboratory\\
Berkeley, California 94720, USA\\
{\tt terning@alvin.lbl.gov}}

\begin{abstract}
We present new theories of dynamical \susy breaking
in which the strong interactions that break \susy also give rise
to composite quarks and leptons with naturally small
Yukawa couplings.
In these models, \susy breaking is communicated directly to the
composite fields without ``messenger'' interactions.
The compositeness scale can be anywhere between
$10\TeV$ and the Planck scale.
These models can naturally solve the \susc
flavor problem, and generically predict
sfermion mass unification independent from gauge
unification.
\end{abstract}

\date{May 7, 1998}

\end{titlepage}

\section{Introduction}
\Susy is arguably the most attractive framework for physics beyond the
standard model, but a truly satisfactory and attractive model for
\susy breaking has yet to emerge.
One reason for dissatisfaction with present models is their
``modular'' structure:
\susy is assumed to be broken in some new sector, and the information
that \susy is broken is communicated to the observable fields via
messenger interactions, which may be either (super)gravity \cite{SUGRA}
or standard-model gauge interactions \cite{OldGaMed,NewGaMed}.%
\footnote{There has recently been important progress in simplifying
models of gauge-mediated \susy breaking \cite{NewerGaMed}.}

While there is in principle nothing wrong with such modular
schemes, it is interesting to ask whether there exist simpler models
in which \susy is broken directly in the observable sector.
An important obstacle in constructing such a model was pointed out
by Dimopoulos and Georgi \cite{DimGeorgi}.
They showed that if one assumes
$(i)$ the gauge group is that of the standard model;
$(ii)$ no higher-dimension operators in the \Kahler potential
of the effective Lagrangian; and
$(iii)$ tree approximation,
then there is always a colored scalar lighter than the down quark.
Any realistic model of \susy breaking must contain important effects
that do not satisfy one of these assumptions.
For example, gravity-mediated models violate $(ii)$, and
gauge-mediated models violate $(iii)$.
The effects that violate $(ii)$ and $(iii)$ are generally smaller than
tree-level renormalizable effects,
but the ``modular'' structure of these models guarantees
that they are the \emph{leading} effects that communicate \susy
breaking to the observable sector.

An interesting way to evade the ``no go'' theorem of Dimopoulos and
Georgi without introducing modular structure is to make the observable
fields composite,
in the sense that they couple to new strong dynamics at a scale $\La$ above
the weak scale.\footnote{Supersymmetric composite models of quarks and
leptons have been previously constructed
\cite{ModComp,NelsonStrass} but these models require separate sectors
for supersymmetry breaking.}
If the strong dynamics also breaks \susy, assumption $(iii)$ will be violated
(and the low-energy theory below the scale $\La$ will violate $(ii)$).
We therefore look for a theory with a single sector that breaks
\susy dynamically and generates composite fermions.

More specifically, we have the following scenario in mind.
Consider a model that breaks \susy by strong interactions at the scale
$\La$, and suppose that the model has an unbroken global symmetry group $G$.
If there are $G^3$ anomalies, the theory will have massless composite
fermions in the low-energy spectrum to match the anomalies \cite{tHooft}.
It is easy to find such models where the standard-model gauge group
$G_{\rm SM}$ can be embedded in $G$, either because there are no
$G_{\rm SM}^3$ anomalies, or because these anomalies are canceled by
``elementary'' states.
In this case, some of the composite fermions will be charged under
$G_{\rm SM}$ and may be identified with quarks and leptons.

If there is no unbroken $\U1_R$ symmetry, standard model gaugino masses
will be generated, suppressed compared to the mass of the composite scalars by
a perturbative loop factor, and one must worry about gaugino masses
being too small.%
\footnote{This killed the models of \Ref{comptry}, which were motivated
by very similar considerations as those described above.}
One possibility is that the composite scalars are heavy enough
that the gauginos are sufficiently heavy despite the loop
suppression factor.
This leads naturally to models with a low compositeness scale and a
superpartner spectrum similar to that of the ``more minimal''
models \cite{CoKaNe}.
Another alternative is to assume that the loop factor is compensated by
a large multiplicity factor.
In fact, in order to generate complete composite generations
the global symmetry must be quite large, so a large multiplicity factor
is hard to avoid.
The large number of states also means that the standard-model
gauge group is far from being asymptotically free, but the models
can still accommodate perturbative gauge coupling unification if the
scale $\La$ of non-perturbative composite
dynamics is near the unification scale.
A large value for $\La$ also helps avoid negative mass-squared terms
for standard-model scalars, as we will explain in the text.

The composite nature of some of the standard-model fermions can also help
in understanding the small Yukawa couplings for the first two generations.
If there are no Yukawa couplings generated by the strong dynamics,
all Yukawa couplings must arise from flavor-dependent higher-dimension
operators in the fundamental theory suppressed by powers of a scale
$M > \La$.
In the low-energy theory, these will become Yukawa couplings suppressed
by powers of $\La / M$.

This class of models makes two interesting generic predictions for
the spectrum of superpartner masses.
First, the gaugino masses will be lighter than the composite scalars.
Second, the composite scalar masses are generated by the strong dynamics,
and are therefore invariant under the global symmetry $G$ at the scale
$\La$.
This means that some or all of the soft masses for the composite
fields unify at the scale $\La$.
If supersymmetry is discovered, this prediction can be
tested if the scalar masses are accurately measured.

\section{New theories of dynamical SUSY breaking}
In this Section, we describe \susc gauge theories that have local minima
with dynamical \susy breaking and composite fermions.
These models are similar in some ways to the models considered in
\Ref{LutyTerning}, but have some features that are more favorable to
the kind of model-building we are interested in.
The models have gauge and flavor symmetry group
\beq
\SU{4} \times \SU{N} \times \left[ \SU{N} \times \U1 \times \U1_R \right]
\eeq
where the group in brackets is a global symmetry group.
The matter content is
\beq\bal
Q &\sim (\fund, \fund)
\times (\hbox{\bf 1}; 1, -\sfrac{N}{4} - 1)~,
\\
L &\sim (\bar{\fund}, \hbox{\bf 1})
\times (\bar{\fund}; -1, \sfrac{N}{4} + 3 - \sfrac{8}{N})~,
\\
\bar{U} &\sim (\hbox{\bf 1}, \bar{\fund})
\times (\fund; 0, \sfrac{8}{N})~,
\\
A &\sim (\hbox{\bf 1}, \asymm\,)
\times (\hbox{\bf 1}; \sfrac{4}{N-2}, 1)~.
\eal\eeq
The theory has a tree-level superpotential
\beq
W = \la L Q \bar{U}.
\eeq
Here $\la$ is a matrix that can be viewed as an adjoint spurion
for the global $[\SU{N}]$ symmetry.
Note that there are 4 $\fund$'s and $N$ $\bar{\fund}$'s under the
global $[\SU{N}]$, so the theory has a nonzero
$\left[ \SU{N} \right]^3$ anomaly for $N \ne 4$.
The analysis of this model is somewhat different depending on whether
$N$ is even or odd, so we consider both possibilities in turn.

\subsection{Odd $N$ Models}
We first consider the case where $N=2n+1$ is odd.
If we include the effects of the tree-level superpotential, the theory
has a classical moduli space that can be parameterized by the gauge
invariants (we indicate the $\left[SU(N)\right]$ quantum numbers)
\beq\bal
L^4 &\sim \bar{\asymmfour} ~,
\quad\hbox{(for\ $N \ge 5$)}
\\
A \bar{U}^2 &\sim \asymm ~,
\\
\bar{U}^{N} &\sim \hbox{\bf 1} ~.
\eal\eeq
with the constraints
\beq
L^4 \cdot \bar{U}^{N} = 0, \quad L^4 \cdot A\bar{U}^2 =0.
\eeq
For $N \ge 5$, the classical moduli space has two branches:
$\avg{\bar{U}^{N}} \ne 0$ with $\avg{L^4} = 0$
(``baryon branch''), and $\avg{L^4} \ne 0$ with $\avg{\bar{U}^{N}} = 0$
(``lepton branch'').
For $N = 3$, only the baryon branch exists.

We first analyze the baryon branch.
In terms of the elementary fields, the \vevs can be written%
%
%
(up to gauge and flavor transformations)
\beq
\avg{Q} = 0,
\quad
\avg{L} = 0
\eeq
\beq
\avg{A} = \sqrt{2} \pmatrix{
a_1 \ep_2 &        &           & 0      \cr
          & \ddots &           & \vdots \cr
          &        & a_n \ep_2 & 0      \cr
0         & \cdots & 0         & 0      \cr},
\quad
\avg{\bar{U}} = \pmatrix{
b_1 \hbox{\bf 1}_2 &        &         & 0      \cr
        & \ddots &         & \vdots \cr
        &        & b_n \hbox{\bf 1}_2 & 0      \cr
0       & \cdots & 0       & b      \cr},
\eeq
where $\hbox{\bf 1}_2$ is the $2 \times 2$ identity matrix,
\beq
\ep_2 \equiv \pmatrix{0 & 1 \cr -1 & 0 \cr},
\eeq
and the \vevs satisfy
\beq
|b_j|^2 = |a_j|^2 + |b|^2,
\quad
j = 1, \ldots, n.
\eeq

We begin by analyzing the theory in the regime where
$b, a_1, \ldots, a_n$ are all nonzero and
large, so that a classical description is valid.
In that case, the gauge group $\SU{N}$ is completely  broken,
while the $\SU{4}$ gauge group remains unbroken.
The fields $Q$ and $L$ get masses $\sim \la \avg{\bar{U}}$.
Integrating out these massive fields gives an effective theory consisting
of $\SU{4}$ super-Yang--Mills theory and some singlets.
$\SU{4}$ gaugino condensation gives rise to a dynamical superpotential%
\footnote{In this Section, we do not include factors of $4\pi$
and $N$ in our estimates for simplicity.
These are included in the numerical estimates we give in the next Section.}
\beq\eql{magicW}
W_{\rm dyn} \propto \det(\la \bar{U})^{1 / 4}.
\eeq
(Alternatively, the anomaly-free $U(1) \times U(1)_R$ symmetries can be
used to show that this is the most general superpotential allowed.)
For large values of $\avg{\bar{U}}$, the \Kahler potential is nearly canonical
in $\bar{U}$, and so the potential slopes toward $\bar{U} = 0$ for
$N \ge 5$.
(For $N = 3$, the theory has a runaway \susc vacuum.)
We are therefore led to analyze the theory near the origin of the
moduli space.%
\footnote{If $b = 0$, $a_1, \ldots, a_n$ large, the analysis is different.
In that case, the $\SU{4}$ gauge group has one light flavor that would run
away if there were no further interactions.
However, the would-be runaway direction is not $D$-flat, so there are no
\susc minima with $b = 0$.}

We analyze the dynamics near the origin of the moduli space assuming
that $\SU{4}$ is weak at the scale $\La_{N}$ where the $\SU{N}$ becomes
strong.
(Note that this includes large values of $N$ where $\SU{4}$ is not
asymptotically free.)
The $\SU{N}$ theory is s-confining, and the effective theory
can be written in terms of the fields \cite{Pouliot}%
\footnote{For a general analysis of s-confining theories, see
\Ref{sconfine}.}
\beq\bal
(Q\bar{U}) &\sim \fund \times \fund ~,
\\
(Q A^n) &\sim \fund \times \hbox{\bf 1} ~,
\\
(Q^3 A^{n - 1}) &\sim \bar{\fund} \times \hbox{\bf 1} ~,
\\
(A \bar{U}^2) &\sim \hbox{\bf 1} \times \asymm ~,
\\
(\bar{U}^{N}) &\sim \hbox{\bf 1} \times \hbox{\bf 1} ~,
\\
L &\sim \bar{\fund} \times \bar{\fund} ~,
\eal\eeq
where we have given the transformation properties under
$\SU{4} \times \left[ \SU{N} \right]$.
The parentheses indicate that these are elementary fields in the
effective Lagrangian with the same quantum numbers as the composite
operators inside the parentheses.
The \Kahler potential is smooth in terms of the effective fields, \eg
\beq
K_{\rm eff} \sim \frac{1}{\La_{N}^{2N - 2}}
\left| (\bar{U}^{N}) \right|^2 + \cdots
\eeq
The effective superpotential is given by the sum of the tree
superpotential and a dynamical superpotential \cite{Pouliot}
\beq\bal
W_{\rm eff} \sim
\frac{1}{\La_{N}^{2N - 1}} &\biggl[
(Q A^n) (Q \bar{U})^3 (A \bar{U}^2)^{n - 1}
+ (Q^3 A^{n - 1}) (Q\bar{U}) (A \bar{U}^2)^n
\\
&\quad
+ (\bar{U}^{N}) (Q A^n) (Q^3 A^{n - 1}) \biggr]
+ \la L (Q\bar{U}).
\eal\eeq
The trilinear term has become a mass term for $L$ and $(Q\bar{U})$;
integrating out these fields gives an $\SU{4}$ gauge theory with 1
flavor $(Q A^n)$, $(Q^3 A^{n - 1})$ and singlets
$(A \bar{U}^2)$, $(\bar{U}^{N})$,
with a trilinear effective superpotential
\beq
W_{\rm eff} \sim \frac{1}{\La_{N}^{2N - 1}}
(\bar{U}^{N}) (Q A^n) (Q^3 A^{n - 1}).
\eeq
If this were a theory of fundamental fields, it would have a runaway vacuum
with $(\bar{U}^{N}) \to \infty$.
This can be described by a superpotential of the form \Eq{magicW}, but
in the regime we are now considering the \Kahler potential is smooth
in terms of the field $(\bar{U}^{N})$.
But if $\avg{\bar{U}}$ is large compared to $\La_{N}$,
we can no longer treat $(\bar{U}^{N})$ as an elementary field;
instead, we must use the analysis above, which shows that the potential
slopes toward $\bar{U} = 0$ for $\avg{\bar{U}} \gg \La_{N}$.
We see that there is no \susc vacuum for either large or small
values of $\bar{U}$ on the baryon branch, so there must be at least a local
\susy-breaking minimum for $\avg{\bar{U}} \sim \La_{N}$.
This is the mechanism for \susy breaking found in the models of
\Refs{LutyTerning,prevbreak}.
Note that there is no unbroken $\U1_R$ symmetry, so that when we
gauge a subgroup of the global $[\SU{N}]$ symmetry, gaugino masses
can be generated.

We see that the baryon branch of this model has two descriptions.
There is a ``Higgs'' description in which the gauge group $\SU{N}$
is broken (valid for large $\avg{\bar{U}}$),
and a ``confining'' description in which $\SU{N}$ confines
(valid near $\avg{\bar{U}} = 0$).
Neither of these descriptions is under control near the local minimum
found above, but both pictures are expected to be a reliable guide to
the qualitative features of the low-energy dynamics \cite{compliment}.
We know that $b \sim \La_{N}$ (using the Higgs description), but
we cannot determine whether $a_1, \ldots, a_n$ are nonzero.
In this paper, we will make the dynamical assumption that
\beq
\avg{A} = 0.
\eeq
This corresponds to the largest possible unbroken global symmetry
\beq\eql{gaugetoglobal}
\SU{N} \times \left[ \SU{N} \right]
\to \left[ \SU{N} \right].
\eeq
This is reasonable, since points of maximal symmetry are generically
stationary points of the energy, but it is an assumption nonetheless.
(The assumption is equivalent to the statement that certain mass-squared
terms in the effective theory are positive.)
With this assumption, we see that the fermionic components of $A$ must remain
massless in order to match the anomalies of the unbroken global
$\left[ \SU{N} \times \U1 \right]$ symmetry.
(In the Higgs description we are using, $A$ is charged under the global
symmetry due to the symmetry breaking in \Eq{gaugetoglobal}.)
If we use the confined description, we find that the fermionic components
of the composite field $(A \bar{U}^2)$ are massless.
In either description, we find that there are massless fermions
transforming as a $\asymm$
under the unbroken $[\SU{N}]$ global symmetry.
Later we will identify some of these fermions with composite quarks and
leptons.

Note that if the dynamical assumption above is false, we can
use this model to construct models of direct gauge-mediated \susy
breaking with composite messengers, along the lines
suggested in \Ref{LutyTerning}.
In this case, we add higher-dimension terms to the superpotential that give
a \susc mass to the composite fields that stabilizes the vacuum
at $\avg{A} = 0$, and gauge a subgroup of the $[\SU{N}]$
global symmetry of this model with
the standard-model gauge group.
The negative \susy-breaking mass-squared terms that result from the
non-perturbative dynamics then induce {\it positive} mass-squared terms
for the squarks and sleptons from gauge loops.
We will not pursue this possibility further in this paper.

In the remainder of this Subsection, we will show that for $N > 5$
there is a runaway \susc vacuum on the lepton branch of the classical
moduli space.
When we consider even $N$, we will find that the story is much the same:
there are three branches of the classical moduli space, and there is a
local \susy-breaking minimum on the ``baryon'' branch whose description
is identical to the one found for $N$ odd, and there are runaway
\susc vacua on the other two branches.
In the remainder of the paper, we will build models assuming that the
universe lives in the false vacuum on the baryon branch.%
\footnote{The rate for tunneling from the false to
one of the \susy vacua is shown to be
negligibly small in Subsection 4.5.}
The rest of this Section is therefore not necessary to understand
the main results of the paper.
The reader interested primarily in model-building is strongly
encouraged to skip to Section 3 at this point.

%

We now analyze the lepton branch of the classical moduli space.
In terms of the elementary fields, the \vevs can be written
(up to gauge and flavor transformations)
\beq
\avg{Q} = 0,
\quad
\avg{L} = \pmatrix{
        & 0      & \cdots & 0      \cr
\ell \hbox{\bf 1}_4 & \vdots &        & \vdots \cr
        & 0      & \cdots & 0      \cr},
\eeq
\beq
\!\!\!\!\!\!\!\!\!
\avg{A} = \pmatrix{
\hbox{\bf 0}_5 &           &        &           \cr
               & a_1 \ep_2 &        &           \cr
               &           & \ddots &           \cr
               &           &        & a_{n - 2} \ep_N \cr}\!,
\ \
\avg{\bar{U}} = \sqrt{2} \pmatrix{
\hbox{\bf 0}_5 &           &        &           \cr
               & a_1 \hbox{\bf 1}_2 &        &           \cr
               &           & \ddots &           \cr
               &           &        & a_{n - 2} \hbox{\bf 1}_2 \cr}\!.
\eeq
We begin by analyzing the theory in the regime where
$\ell, a_1, \ldots, a_{n - 2}$ are all nonzero and large,
so that a classical description is valid.
In that case, the gauge group $\SU{4}$ is completely broken, and
$\SU{N}$ is broken down to $\SU{5}$.
$\avg{L} \ne 0$ gives mass to all of the $Q$'s and 4 flavors of $\bar{U}$'s,
and most of the components of $A$ and $\bar{U}$ are eaten.
The effective $\SU{5}$ gauge theory has matter content
$\bar{\fund} \oplus \asymm$
plus singlets, with no superpotential.
If this were a theory of fundamental fields, this would break \susy
\cite{fivebarten}, giving a vacuum energy proportional to
$\La_{5,{\rm eff}}^4$, where $\La_{5,{\rm eff}}$ is determined by 1-loop
matching to be
\beq
\La_{5,{\rm eff}} = \La_{N}^{(4N + 1)/13}
\ell^{4/13} (a_1 \cdots a_{n-2})^{-(4N - 8) / (13(n - 2))}.
\eeq
For $N > 5$, the vacuum energy goes to zero as the $a$'s go to infinity,
and there are runaway vacua on the lepton branch.
For $N = 5$, the classical constraints force $\avg{A} = 0$, and there
are no runaway directions;
in this case \susy is broken on the lepton branch as well as the
baryon branch.

For $N > 5$, we could lift the runaway directions by adding
higher-dimension terms to the superpotential (see \cite{ADStoo}).
However, these will partially break the global symmetry, and can be shown
to have lower energy than the local minima on the baryon branch.

\subsection{Even $N$ Models}
We now consider the model for even $N = 2n$.
The analysis closely parallels that of the odd $N$ models, and the reader
interested mainly in our models is encouraged to skip to Section 3.
The classical moduli space can be parameterized by (again indicating the
global $[\SU{N}]$ quantum numbers)
\beq\bal
L^4 &\sim \bar{\asymmfour} ~,
\qquad\hbox{(for\ $N \ge 4$)}
\\
A \bar{U}^2 &\sim \asymm ~,
\\
Q^4 A^{n - 2} &\sim \hbox{\bf 1} ~,
\\
\bar{U}^{N} &\sim \hbox{\bf 1} ~,
\\
A^n &\sim \hbox{\bf 1} ~,
\eal\eeq
with the constraints
\beq\bal
L^4 \cdot \bar{U}^{N} &= 0,
\quad
L^4 \cdot A\bar{U}^2 = 0,
\quad
L^4 \cdot Q^4 A^{n - 2} = 0,
\quad
\bar{U}^{N} \cdot Q^4 A^{n - 2} = 0.
\eal\eeq
This moduli space has three branches:
$\avg{L^4} \ne 0$,
$\avg{\bar{U}^{N}}, \avg{Q^4 A^{n - 2}} = 0$
(``lepton branch''),
$\avg{\bar{U}^{N}} \ne 0$,
$\avg{L^4}, \avg{Q^4 A^{n - 2}} = 0$
(``baryon branch''),
and $\avg{Q^4 A^{n - 2}} \ne 0$,
$\avg{L^4}, \avg{\bar{U}^{N}} = 0$
(``mixed branch'').

We first analyze the baryon branch.
On this branch, the \vevs can be written
(up to gauge and flavor transformations)
\beq
\avg{L} = 0,
\quad
\avg{Q} = 0,
\eeq
\beq
\avg{A} = \pmatrix{a_1 \ep_2 & & \cr & \ddots & \cr & & a_n \ep_2 \cr},
\quad
\avg{\bar{U}} = \sqrt{2} \pmatrix{b_1 \hbox{\bf 1}_2 & & \cr & \ddots & \cr
& & b_n \hbox{\bf 1} \cr},
\eeq
where
\beq
|a_j|^2 - |b_j^2| = c,
\quad j = 1, \ldots, n.
\eeq

We begin by analyzing the theory in the region of moduli space where
$a_1, \ldots a_n$ are all nonzero and large, so that a classical
description is valid.
In that case, the gauge group $\SU{N}$ is completely broken, while the
$\SU{4}$ gauge group remains unbroken.
The fields $Q$ and $L$ get masses $\sim \la\avg{\bar{U}}$, and the low-energy
theory is $\SU{4}$ super-Yang--Mills with singlets.
Gaugino condensation in this theory gives rise to a dynamical superpotential
\beq
W_{\rm dyn} \propto \det(\la \bar{U})^{1 / 4}.
\eeq
For large values of $\avg{\bar{U}}$, the \Kahler potential is nearly canonical
in $\bar{U}$, and so the potential slopes toward $\bar{U} = 0$ for $N > 4$.
For $N = 2$, there is a runaway \susc vacuum.
For $N = 4$, the superpotential is linear in $\bar{U}$, and the location
of the true vacuum depends on the form of the \Kahler potential.
For large values of $\vu$, the \Kahler potential can be computed in
perturbation theory, and one finds that 1-loop corrections involving the
Yukawa coupling $\la$ tend to push the the vacuum away from the origin,
while 1-loop corrections involving the gauge couplings have the opposite
sign.
These effects can give rise to a local minimum for large values of
$\vu$ for a range of parameters.
(This is the inverted hierarchy mechanism \cite{inverted}.)
For any $N \ge 4$, we see that there is no \susc vacuum for large
$\vu$, and we are led to analyze the theory near the origin of the
moduli space.

We now analyze the dynamics near the origin of the moduli space assuming
that $\La_{N} \gg \La_4$.
The $\SU{N}$ theory is s-confining, and the effective theory can be written in
terms of the fields \cite{Pouliot}
\beq\bal
(Q\bar{U}) &\sim \fund \times \fund ~,
\\
(A^n) &\sim \hbox{\bf 1} \times \hbox{\bf 1} ~,
\\
(Q^2 A^{n - 1}) &\sim \asymm
\times \hbox{\bf 1} ~,
\\
(Q^4 A^{n - 2}) &\sim \hbox{\bf 1} \times \hbox{\bf 1} ~,
\\
(A \bar{U}^2) &\sim \hbox{\bf 1} \times \asymm ~,
\\
(\bar{U}^{N}) &\sim \hbox{\bf 1} \times \hbox{\bf 1} ~,
\\
L &\sim \bar{\fund} \times \bar{\fund} ~,
\eal\eeq
where we have given the transformation properties under
$\SU{4} \times \left[ \SU{N} \right]$.
The superpotential is given by the sum of the tree superpotential and
a dynamical superpotential \cite{Pouliot}.
The tree-level superpotential turns into a mass term for $L$ and
$(Q\bar{U})$.
Integrating out these states gives an effective theory with gauge
group $\SU{4}$, a field $(Q^2 A^{n - 1}) \sim \asymm\,$, and singlets,
with effective superpotential
\beq\bal
W_{\rm eff} \sim
\frac{1}{\La_{N}^{2N - 1}} &\biggl[
(Q^4 A^{n - 2}) (A \bar{U}^2)^n
+ (\bar{U}^{N}) (A^n) (Q^4 A^{n - 2})
\\
&\quad
+ (\bar{U}^{N}) (Q^2 A^{n - 1})^2 \biggr].
\eal\eeq
For $\avg{\bar{U}^{N}} \ne 0$, $(Q^2 A^{n - 1})$ is massive and
$\SU{4}$ gaugino condensation pushes $(\bar{U}^{N})$ away from the
origin.
If this were a theory of fundamental fields, it would have a runaway vacuum
with $(\bar{U}^{N}) \to \infty$, but this description breaks down for
large values of $\avg{\bar{U}}$.
We see that this theory has a local \susy-breaking minimum on the baryon
branch through a mechanism identical to that in the odd $N$ case.
As before, we make the dynamical assumption that
\beq
\avg{A} = 0,
\eeq
so that there is an unbroken $[\SU{N}]$ global symmetry, and the theory
has massless composite fermions transforming as a $\asymm$
under $\SU{N}$.

We now turn to the lepton branch.
On this branch, the \vevs are
\beq
\avg{L} = \pmatrix{& 0 & \cdots & 0 \cr
\ell \hbox{\bf 1}_4 & \vdots & & \vdots \cr
& 0 & \cdots & 0 \cr},
\quad
\avg{Q} = 0,
\eeq
\beq
\!\!\!\!\!\!\!\!
\avg{A} \! = \! \pmatrix{a \epsilon_2 \!\! & & & & \cr
& a \epsilon_2 \!\! & & & \cr
& & a_1 \epsilon_2 \!\! & & \cr
& & & \ddots & \cr
& & & & a_{n-2} \ep_2 \!\! }\!,
\
\avg{\bar{U}} \! = \! \sqrt{2} \pmatrix{\hbox{\bf 0}_4 & & & \cr
& b_1 \hbox{\bf 1}_2 & & \cr
& & \ddots & \cr
& & & b_{n - 2} \hbox{\bf 1}_2 \cr}\!.
\eeq
with
\beq
|a|^2 = |a_j|^2 - |b_j|^2,
\quad j = 1, \ldots, n - 2.
\eeq
We analyze the theory in the region of moduli space where
$\ell, a, a_1, \ldots, a_{n - 2}$ are all nonzero and large.
In that case, the $\SU{4}$ gauge group is completely broken,
and the $\SU{N}$ gauge group is broken down to $\Sp{4}$.
After taking into account the effects of the superpotential and
the eaten fields, there are no charged fields under the unbroken
$\Sp{4}$.
Gaugino condensation in $\Sp{4}$ then pushes $a, a_1, \ldots, a_{n-2}$
away from the origin \cite{ADStoo}, and so there is a runaway
\susc vacuum in this branch.

Finally, we analyze the mixed branch.
On this branch, the \vevs are
\beq
\avg{L} = 0,
\quad
\avg{Q} = \sqrt{2} \pmatrix{& 0 & \cdots & 0 \cr
q \hbox{\bf 1}_4 & \vdots & & \vdots \cr
& 0 & \cdots & 0 \cr}\!,
\eeq
\beq
\!\!\!\!\!\!\!\!
\avg{A} \! = \! \pmatrix{a \epsilon_2 \!\! & & & & \cr
& a \epsilon_2 \!\! & & & \cr
& & a_1 \ep_2 \!\! & & \cr
& & & \ddots & \cr
& & & & a_{n - 2} \ep_2 \!\! \cr}\!,
\
\avg{\bar{U}} \! = \! \sqrt{2} \pmatrix{\hbox{\bf 0}_4 & & & \cr
& b_1 \hbox{\bf 1}_2 & & \cr
& & \ddots & \cr
& & & b_{n - 2} \hbox{\bf 1}_2 \cr},
\eeq
where
\beq
|a|^2 + |q|^2 = |a_j|^2 - |b_j|^2,
\quad j = 1, \ldots, n - 2.
\eeq
As on the lepton branch, the low-energy theory is a pure
$Sp(4)$ gauge symmetry,
and gaugino condensation pushes $a,a_j$ away from the origin, and so
there are additional runaway vacua on this branch.

\section{Numerical Estimates}
We now consider the numerical estimates of various quantities of
interest in these models.
The models are non-calculable, but we can make estimates using
dimensional analysis, and also keep track of factors of $4\pi$ and $N$,
which are potentially large.
Neither the ``Higgs'' nor the ``confining'' descriptions of these theories
is weakly coupled in the local \susy-breaking vacuum we consider.
We find it simplest to make estimates using the ``Higgs'' description
that uses the elementary fields of the theory.
We estimate the size of various effects by assuming that loop corrections
are the same size as leading effects in perturbation theory.
This is the philosophy of ``na\"\i ve dimensional
analysis" \cite{fourpi,fourpiSUSY}.

We use these considerations to argue that the strong dynamics preserves
an approximate $[\SU{N}]$ flavor symmetry even if the
Yukawa matrix $\la$ in the tree-level superpotential is completely
arbitrary.%
\footnote{It is a consequence of our dynamical assumption that
$\avg{A} = 0$, i.e. that the flavor symmetry is not spontaneously broken
by the strong dynamics in the limit where $\la$ is proportional to the
identity.}
This is important for naturally suppressing flavor-changing neutral
currents in the models we construct below.%
\footnote{We thank M. Schmaltz for emphasizing this point.}
First of all, the dynamical superpotential \Eq{magicW} depends only
on $\det(\la)$, and so has no flavor dependence.
This means that all flavor dependence appears in the effective \Kahler
potential.
In the Higgs description,
the $\la$ dependence in the \Kahler potential comes from diagrams
with $\la$ vertices, and through the Dirac mass matrix of $Q$ and
$L$, which is proportional to $\la$.
Diagrams with $\la$ vertices are suppressed by $\la^2 / (16\pi^2)$,
so these give only small flavor violation.
Internal $Q$ and $L$ loops without $\la$ vertices do not contribute
to flavor violation because they always involve traces of the mass
matrix.
(We are not interested in diagrams with external $Q$ and $L$ lines
because the only light matter states correspond to $\tr\bar{U}$
and $A$; see below.)
This shows that the flavor symmetry is preserved up to corrections
of order $\la^2 / (16\pi^2) \lsim 10^{-2}$.

We now estimate $\avg{\bar{U}}$.
Na\"\i ve dimensional analysis tells us that $\avg{\bar{U}}$ must
be close to the value for which perturbation theory breaks down.
The $\SU{N}$ gauge dynamics becomes strong at the scale $\La_N$,
where
\beq
g_N(\mu \sim \La_N) \sim \frac{4\pi}{\sqrt{N}}.
\eeq
The perturbative description breaks down when the massive gauge bosons
(and the states that get a mass due to the $\SU{N}$ $D$-term potential)
have masses of order $\La_N$, which gives
\beq
\avg{\bar{U}} \sim \La_N \frac{\sqrt{N}}{4\pi} \hbox{\bf 1}_N.
\eeq
The $F$ component of $\bar{U}$ is estimated to be%
\footnote{For a discussion of the factors of $4\pi$ in $W_{\rm eff}$,
see \Ref{fourpiSUSY}.}
\beq
\avg{F_{\bar{U}}} \sim \left\langle
\frac{\partial W_{\rm eff}}{\partial \bar{U}} \right\rangle
\sim \frac{1}{4\pi} \frac{\sqrt{N}}{4}
\left( \det(\sqrt{N}\la) \right)^{1/4}
\La_4^{3 - N/4} \La_N^{N/4 - 1} \hbox{\bf 1}_N.
\eeq
This shows that as long as $\SU{4}$ is weak at the scale where $\SU{N}$
becomes strong, we have $\avg{F_{\bar{U}}} \ll \avg{\bar{U}}^2$.
If $N < 12$, $\SU{4}$ is asymptotically free and the condition for
$\SU{4}$ to be weak at the scale $\La_N$ is $\La_4 \ll \La_N$.
For $N \ge 12$, $\La_4$ is the ultraviolet Landau pole of $\SU{4}$,
and so the condition that $\SU{4}$ is weak at $\La_N$ is
$\La_4 \gg \La_N$.
As a consequence of our dynamical assumption,
both $\avg{\bar{U}}$ and $\avg{F_{\bar{U}}}$ are
proportional to the $N \times N$
unit matrix, so that the $\SU{N} \times \left[ \SU{N} \right]$
symmetry is broken down to a global $[\SU{N}]$.

The superpotential gives a \susc mass to the fields $Q$ and $L$ of order
\beq
m_{Q,L} \sim \la \avg{\bar{U}}
\sim \frac{\sqrt{N} \la}{4\pi} \La_N.
\eeq
(This mass does not become large compared to $\La_N$
for large $N$ because the Yukawa coupling must be $\la \sim 1/\sqrt{N}$
in order to have a good large-$N$ limit.)
There are also \susy breaking $B$-type mass terms of order
$\avg{F_{\bar{U}}}$.
Below the scale $m_{Q,L}$, the only light fields are the $\SU{4}$ gauge
bosons, $\tr\bar{U}$ and $A$.
(In the confined description, these fields correspond to
$(\bar{U}^N)$ and $(A\bar{U}^2)$, respectively.)
The field $\tr\bar{U}$ is a singlet,
and $A$ transforms as a $\asymm$ under the
unbroken $[\SU{N}]$ global symmetry.
The scalar and fermion components of $\tr\bar{U}$ get masses of
order
\beq
m_{\tr\bar{U}} \sim \left\langle
\frac{\partial^2 W_{\rm eff}}{\partial\bar{U}^2} \right\rangle
\sim \frac{\avg{F_{\bar{U}}}}{\avg{\bar{U}}}
\equiv M_{\rm comp}.
\eeq
We will see that the scale $M_{\rm comp}$ sets the scale for all
\susy breaking masses in this model.

The scalar components of the field $A$ receives (strong \SU{N}
gauge-mediated) loop
contributions both from the \susy breaking in the $Q,L$ spectrum and
from the induced \susy breaking in the fields at the scale $\La_N$.
These contributions can be most easily estimated using the method
of Giudice and Rattazzi \cite{GiudRatt}.%
\footnote{This method can be extended to all orders in perturbation
theory \cite{SonOfGiudRatt}.}
In this method, one computes the wavefunction renormalization factor
$Z_A$ as a function of the threshold $m$ where heavy states are integrated
out, and then makes the replacement
$m \to 4\pi \sqrt{\bar{U}^\dagger \bar{U}} / \sqrt{N}$ to find
the dependence on $\avg{\bar{U}}$ and $\avg{F_{\bar{U}}}$ to
leading order in $\avg{F_{\bar{U}}}/\avg{\bar{U}}$.
The $A$ scalar mass is then obtained from the $\theta^2 \bar{\theta}^2$
component of $\ln Z_{A}$.
The quantity $\ln Z_{A}$ satisfies a renormalization group equation
\beq
\mu \frac{d}{d\mu} \ln Z_{A} = f\left(\frac{N g_N^2}{16\pi^2}\right),
\eeq
where $f$ is a function with no large parameters.
(Note that there are $N$ ``flavors'' of $\bar{U}$, so loops of $\bar{U}$
fields are not suppressed for large $N$.)
Since $N g_N^2 / (16\pi^2) \sim 1$, we obtain simply
\beq
m^2_{\phi_A} \sim \left(
\frac{\avg{F_{\bar{U}}}}{\avg{\bar{U}}} \right)^2
= M_{\rm comp}^2.
\eeq
If we identify the composite fermions with quarks and leptons, this
gives the mass of the corresponding scalar superpartners.

We now assume that the standard-model gauge group is embedded into
the $[\SU{N}]$ global symmetry and estimate the standard-model gaugino
and elementary scalar masses.
We can compute these using the method of \Ref{GiudRatt}, or by simply
estimating the corresponding perturbative diagrams.
There are of order $N$ messengers, so we obtain
\beq
m_{\la_{\rm SM}} \sim N \frac{g^2_{\rm SM}}{16\pi^2} M_{\rm comp}
\eeq
for the standard-model gaugino masses.
In addition, the scalars will receive a gauge-mediated contribution
\beq
\de m^2_{\phi,{\rm gauge\,med}}
\sim N \left( \frac{g^2_{\rm SM}}{16\pi^2} \right)^2 M_{\rm comp}^2.
\eeq
For the composite fields, this is a small correction;
for the elementary fields, this is the dominant contribution to the
scalar mass.
(We will see below that there is also a flavor-dependent contribution
to the scalar masses that can be comparable.)

In the models we consider, there is a scale of new physics $M$ that
is not far above the scale $\La_N$.
In the effective theory at the scale $\La_N$, there will therefore
be higher-dimension operators suppressed by powers of $1/M$.
For example, the following terms in the Lagrangian are compatible with
all symmetries:
\beq\eql{softmassbreak}
\de\scr{L} \sim \myint d^4\th \left[
\frac{c_1}{M^2} \tr( \bar{U}^\dagger \bar{U} ) A^\dagger A
+ \frac{c_2}{M^2} \tr( \bar{U}^\dagger \bar{U} ) \Phi^\dagger \Phi
\right],
\eeq
where $\Phi$ is an elementary quark or lepton field.
In the ``Higgs'' picture we are using, we can estimate the terms in the
effective Lagrangian for the composite fields by simply replacing $\bar{U}$
by its \vev.%
\footnote{When expressed in terms of the scale $\La_N$, this gives results
with $4\pi$ dependence in agreement with a
``confined'' description \cite{fourpiSUSY}.}
We therefore obtain an additional contribution to the elementary and
composite scalar masses of order
\beq\eql{seec}
\de m_{\phi,{\rm new\,phys}}^2 \sim \frac{c_{1,2} N \avg{F_{\bar{U}}}^2}{M^2}
\sim \frac{c_{1,2} N \avg{\bar{U}}^2}{M^2}
M_{\rm comp}^2.
\eeq
On general grounds, we might expect $c_{1,2} \sim 1$;
alternatively, if the model has a good large-$N$ limit with $M$ held
fixed, we expect $c_{1,2} \sim 1/N$.

There are additional higher-dimension operators in the models we
construct.
We can easily estimate their effects on the composite fields in the
Higgs description by simply replacing $\bar{U}$ by appropriate
scalar or $F$-component \vevs.

\section{Composite Quarks and Leptons}
We now build models of composite quarks and leptons using the models
analyzed above as building blocks.
Because the Yukawa couplings arise from high-dimension operators,
they are naturally small compared to unity.
This means that the top quark cannot be composite in the models
we construct.%
\footnote{It would be interesting to find \susy-breaking models where
the top-quark Yukawa coupling arises as a term in a dynamical superpotential.
In that case, the top-quark Yukawa coupling is of order $4\pi$ at
the compositeness scale, and runs down to a quasi-fixed point
value at the weak scale \cite{fixed}.
The top-quark Yukawa coupling arises in this way in the models of
Nelson and Strassler \cite{NelsonStrass}, but the composite dynamics
does not break \susy in these models.}
In models of this type, the masses of the gauginos and elementary
scalars are suppressed by a loop factor compared to the composite scalar
masses:
\beq
\frac{m_{\la_{\rm SM}}}{M_{\rm comp}}
\sim \frac{N g_{\rm SM}^2}{16\pi^2}.
\eeq
If $N$ is not large, then this can be realistic only if the composite
scalars are very heavy.
As we will explain below, this leads naturally to models with a low
compositeness scale.
On the other hand, we can consider models where the loop suppression is
overcome by the large multiplicity factor $N$, allowing models with
a high compositeness scale.%
\footnote{We do not consider the possibility that the gauginos may
be ultra-light \cite{Glennys}.}

\subsection{Embedding the Standard Model}
Before turning to the models, we discuss some aspects of
embedding the standard model gauge group
into the global $[\SU{N}]$ symmetry.
Because we want to preserve perturbative unification, we will consider
only embeddings where the preons fall into complete $\SU{5}_{\rm SM}$
multiplets, even if only the standard-model subgroup is gauged.%
\footnote{The interesting possibility that the preons fall into complete
representations of the ``trinification'' group $\SU{3}^3 / Z_3$ will not
be explored in this paper.}

Because our models generate composite states transforming as a
$\asymm$ of a global $[\SU{N}]$ symmetry,
it is tempting to generate a $\bb{10}$ of $\SU{5}_{\rm SM}$ from
the antisymmetric product $(\bb{5} \otimes \bb{5})_{\rm asymm}$.
However, it is easy to see that there is no way of assigning
baryon number to the preons to obtain the correct baryon numbers
for the states of the
the composite $\bb{10}$.%
\footnote{It is possible to obtain a ${\bf 10}$ from the antisymmetric
product of two \emph{different} ${\bf 5}$'s, but this leads to rather
uneconomical models.}
Since baryon number is not a good quantum number of the strong
dynamics, we expect baryon-number violating operators suppressed by
powers of $\La_N$ in the low-energy theory, so this kind of embedding
cannot be used in models where the compositeness scale is below the
grand-unified theory (GUT) scale.
It is not hard to construct baryon-number conserving as well as
baryon-number violating embeddings, and we will consider both types
below.

The first embedding we consider is based on the model with $N = 11$.
$\SU{5}_{\rm SM}$ is embedded into $[\SU{11}]$ so that the
$\fund = {\bf 11}$ representation decomposes as
\beq
\fund \to \bb{5} \oplus \mybar{\bb{5}} \oplus \bb{1}.
\eeq
The composite states then decompose under $\SU{5}_{\rm SM}$ as
\beq
\asymm \to \bb{10} \oplus \mybar{\bb{5}} \oplus \bb{1}
\oplus \left[ \bb{24} \oplus \mybar{\bb{10}} \oplus \bb{5} \right].
\eeq
The composite states include a complete generation (including a
right-handed neutrino), together with the exotic states in
square brackets.
Baryon number is violated at the scale $\La_N$.
We can remove the unwanted exotic states by adding an additional
elementary generation $\bb{10} \oplus \mybar{\bb{5}}$ to the theory
and including higher-dimension operators of the form
\beq
\de\scr{L}_{\rm eff} \sim \myint d^2\th \left[
\frac{1}{M} (A \bar{U}^2)_{\bf 5}
X_{\vphantom{\mybar{\bf 5}}\mybar{\bf 5}}
+ \frac{1}{M} (A \bar{U}^2)_{\vphantom{\mybar{\bf 5}}\mybar{\bf 10}}
X_{\bf 10}
+ \frac{1}{M^3} (A \bar{U}^2)_{\bf 24}^2
\right] + \hc,
\eeq
which gives rise to masses
\beq
m_{\mybar{\bf 5}, {\bf 10}} \sim \frac{\vu^2}{M},
\quad
m_{\bf 24} \sim \frac{\vu^4}{M^3}.
\eeq
One can obtain a model with two composite generations by
considering a model with gauge group
$\left[ \SU{4} \times \SU{11} \right]^2/Z_2$.
This may not be unnatural, since whatever explains the replication of
families may also give rise to a replicated group structure.

A simple way to conserve baryon number is to have only composite
$\mybar{\bb{5}}$'s.
The simplest such model is based on $N = 5 + k = 7$ with $\SU{5}_{\rm SM}$
acting on the preons as
\beq
\fund \to \mybar{\bb{5}} \oplus (k \times \bb{1}).
\eeq
The composite states decompose as
\beq
\asymm \to (k \times \mybar{\bb{5}}) \oplus \left[
\mybar{\bb{10}}
\oplus \left( \sfrac{k(k - 1)}{2} \times \bb{1} \right) \right].
\eeq
This gives rise to $k = 2$ composite $\mybar{\bb{5}}$'s and some unwanted
states that can be eliminated by adding higher-dimension operators
similar to those described above.

Finally, we consider a more elegant embedding that naturally
replicates generations and conserves baryon number.
We consider the theory with $N = 15 + k = 18$, with $\SU{5}_{\rm SM}$ acting
on the preons as
\beq
\fund \to \bb{10} \oplus \mybar{\bb{5}} \oplus (k \times \bb{1}).
\eeq
Then the composite states decompose as
\beq
\asymm \to (k \times \bb{10}) \oplus (k \times \mybar{\bb{5}}) \oplus
\left[ {\bf 45} \oplus \mybar{{\bf 45}} \oplus
\mybar{{\bf 10}} \oplus \bb{5}
\oplus \left( \sfrac{k(k - 1)}{2} \times {\bf 1} \right) \right].
\eeq
If we now write down the most general superpotential involving
the composite states, we will generate Dirac masses marrying
$\bb{45}$ and $\mybar{\bb{45}}$, as well as marrying one of the
composite generations with the antigeneration, leaving us with
$(k -1)$ complete composite generations.

We now address the question of Yukawa couplings.
Yukawa couplings involving the composite fermions must arise from
higher-dimension operators in the fundamental theory.
We therefore assume that the new physics at the scale $M$ induces
terms in Lagrangian such as
\beq\eql{YukOp}
\de\scr{L} \sim \myint d^2\th \left[
\frac{b_1}{M^4} (A \bar{U}^2)_{\mybar{\bf 5}}
(A \bar{U}^2)_{\vphantom{\mybar{\bf 5}}\bf 10} H
+ \frac{b_2}{M^2} (A \bar{U}^2)_{\mybar{\bf 5}}
\Phi_{\vphantom{\mybar{\bf 5}}\bf 10} H
+ \cdots \right] + \hc
\eeq
where $H$ is a fundamental Higgs field and $\Phi$ is a fundamental
matter field.
This gives Yukawa couplings to the composite fields
\beq
\de\scr{L}_{\rm eff} \sim \myint d^2\th \left[
\frac{b_1 \avg{\bar{U}}^4}{M^4} A_{\mybar{\bf 5}}
A_{\vphantom{\mybar{\bf 5}}\bf 10} H
+ \frac{b_2 \avg{\bar{U}}^2}{M^2} A_{\mybar{\bf 5}}
\Phi_{\vphantom{\mybar{\bf 5}}\bf 10} H
+ \cdots \right] +\hc
\eeq
Note that if the second generation quarks are to be composite, we require
$\avg{\bar{U}} / M \sim \sfrac{1}{3}$,
so the scale of new physics is not far above the scale of strong
dynamics.
This problem appears particularly worrisome if we note that the scale
$\avg{\bar{U}}$ is smaller than $\La_N$ (the scale of strong dynamics)
for moderate $N$.
However, the example of the charm quark in QCD suggests that it is
not absurd to integrate out particles with masses near the scale $\La_N$.
(The charm quark mass and the scale $\La$ in QCD are both near 1~GeV.)

Note that, in all of the above models, an approximate flavor symmetry
of the strong dynamics (the $Z_2$ in the $N = 11$ theory, $\SU{k}$
in the $N = 5 + k$ and $N = 15 + k$ theories)
guarantees equal soft masses for all the composite states.
While this is somewhat artificial in the $N = 11$ case, it
is quite natural in the $N = 5 + k$ and $15 + k$ cases.
In particular, in the $N = 15 + k$ case, {\it all} soft masses for the
first two generation scalars are degenerate at leading order.
Of course, the flavor physics responsible for
generating the correct pattern of Yukawa couplings must distinguish between
the first two generations and will necessarily break the flavor
symmetry of the strong dynamics.%
\footnote{Flavor violation in the the $\la$ matrix does not break
the chiral symmetries acting on the composite quarks and leptons,
and therefore does not give rise to Yukawa couplings.}
The corrections to the soft masses induced by this flavor physics
are model-dependent, but are at least suppressed by the same small
parameters that control the small Yukawa couplings for the light
generations.
We will see that this suppression is already sufficient for marginal
consistency with flavor-changing neutral current (FCNC) constraints,
so the \susc flavor problem is very mild in these models.

\subsection{Low-scale Composite Models}
If the multiplicity factor $N$ is not large, then the composite scalars
must have masses of order $10\TeV$ or more in order to have gaugino
and elementary squark and slepton masses of order $100\GeV$.
In this case, there are negative 2-loop contributions to the elementary
scalar mass-squared from the composite scalar masses \cite{NimaHitoshi}.
These contributions are dangerous because they are enhanced by
$\ln(\La_N / M_{\rm comp})$ compared to the usual gauge-mediated
contributions.
To avoid these, we must require that $\La_N$ is not far above
$M_{\rm comp} \sim 10$--$100 \TeV$.
Independently of these considerations, we are interested in
the possibility of a low compositeness scale because it holds out the
possibility of rich phenomenology.

One possibility is to use the $N = 5 + k = 7$ model, which gives rise to
the composite states
$\mybar{\bb{10}} \oplus (2 \times \mybar{\bb{5}}) \oplus \bb{1}$.
We identify the two composite $\mybar{\bb{5}}$ fermions with quarks
and leptons, and eliminate the unwanted composite fermions by
combining them with elementary fields transforming as
$\bb{10} \oplus \bb{1}$.
In order to obtain sufficiently heavy masses for the elementary squarks
and sleptons, we take the mass of the composite scalars to be in the
$10$--$100\TeV$ range.
For purposes of running the standard-model gauge couplings,
this model adds an equivalent of $6 \times (\bb{5} \oplus \mybar{\bb{5}})$
to the theory above the scale $\La_7$ of the strong $\SU{7}$ dynamics,
and so it is marginally compatible with unification if
$\La_7 \gsim 200\TeV$.

If we assume that the Yukawa couplings are generated by new physics
at a scale $M$ from operators of the form \Eq{YukOp}, we find that
in order to generate Yukawa couplings of order $10^{-3}$
(for the composite $s$ and $\mu$), we require $\vu / M \sim 3
\times 10^{-2}$.
This gives an explanation of the smallness of the down-type Yukawa
couplings of the first two generations, but it does not explain why
the up-type Yukawa couplings are also small.

We now discuss FCNC's in this model.
Note that there is a global $\SU{2}$ acting on the $SU(5)_{\rm SM}$
singlet preons
in this model, which becomes a $\SU{2}$ flavor symmetry acting on the
composite $\bb{5}$'s in the low-energy theory.
We can therefore envision that the flavor breaking in the preon
theory has a GIM mechanism acting on the first two generations
that would align the flavor structure in the scalar and fermion
sectors \cite{U(2),NS,techniGIM}.
In the absence of such a mechanism, this model has FCNC's.
Because the up-squarks are elementary, their mass arises dominantly
from gauge-mediation, and this is not large enough to naturally
suppress FCNC's.
For example,
\beq
\frac{\de m_{\tilde{u} \tilde{c}}^2}{m_{\tilde{u}}^2}
\sim \left( \frac{\vu}{M} \right)^2
\left( \frac{g_3^2}{16\pi^2} \right)^{-2}
\sim 1,
\eeq
where we use $\vu^2 / M^2 \sim y_{uc} \sim \sqrt{y_u y_c}$.
This is incompatible with the bound from $D$--$\bar{D}$ mixing,
which requires
$\de m_{\tilde{u} \tilde{c}}^2 / m_{\tilde{u}}^2 \lsim 10^{-2}$.
There are also problems with $K$--$\bar{K}$ mixing.

We next consider a model based on the $N = 15 + k = 18$ embedding
described above.
With such a large value for $N$, it may not be necessary to have a low
value for $\vu$ to avoid negative third generation scalar masses, but
we can consider the possibility of a low compositeness
scale nonetheless.
This model produces 2 complete composite generations of quarks
and leptons, but contains a large number of fields charged under
the standard model gauge group above the scale $\La_{18}$ of the
strong $\SU{18}$ dynamics.
The standard-model gauge couplings have a Landau pole at a few
times $\La_{18}$ in this model; so it is certainly not
compatible with perturbative unification.
Since the Landau pole is so close to $\La_{18}$ is not clear that
this model makes sense as an effective theory at the scale $\La_{18}$.
However, the strong dynamics at the Landau scale may have an
interpretation in terms of a dual theory \cite{dual}, and we expect
such a theory to behave qualitatively the same as what we find here.
We can also hope that models with a more favorable group-theory
structure will be found.

The problems with perturbative unification lead us naturally to
consider high-scale models with large values of $\La_N$.
The high-scale and low-scale models with two composite generations
have a similar phenomenology,
and we will discuss this after we have introduced the high-scale
models.

\subsection{High-scale Composite Models}
We now discuss the possibility that the compositeness scale $\Lambda_N$ is
near or above $M_{\rm GUT} \sim 10^{16}\GeV$,
allowing perturbative unification even if $N$ is large.
If the scale $\La_N$ is large, we must address the dangerous
negative contributions to the third generation scalar masses coming
from the scalars of the first two generations \cite{NimaHitoshi}.
These arise from the renormalization group equations
\beq
\mu \frac{d m_3^2}{d\mu} = \frac{8 g^2}{16\pi^2} C_2 \left[
\frac{3 g^2}{16\pi^2} m_{1,2}^2 - m_\la^2 \right],
\eeq
where we have assumed that one gauge group dominates and
specialized to the case of two composite generations.
(Here, $m_3$ is the third-generation scalar mass, $m_{1,2}$ are the
scalar masses of the first two generations, and $m_\la$ is 
the gaugino mass.
$C_2$ is the quadratic Casimir, with the $\U1_Y$ generator in
$\SU{5}$ normalization.)
We see that the contribution to the gaugino mass dominates provided that
\beq
m_\la \gsim \frac{m_{1,2}}{10},
\eeq
which agrees with the detailed analysis of \Ref{NimaHitoshi}.
This condition is plausibly satisfied in our models if $N \gsim 10$.

Since dimension-6 $B$-violating operators suppressed by such high scales
are safe, we consider both the $B$-violating ``squared" $N = 11$ as well
as the $B$-conserving $N = 18$ theories.
Both of these theories give rise to two complete composite
generations.

It is believed that new physics at the Planck scale will give rise to
higher-dimension operators suppressed by the reduced Planck scale
$M_* \sim 10^{18}\GeV$.
It is therefore natural to consider the possibility that it is these
effects that give rise to the higher-dimension operators that are
required to make the theory realistic, and identify $M = M_*$.
In this case, $\La_N$ will be \emph{above} $M_{\rm GUT}$, and
even those extra charged states that become
massive due to higher dimension operators are
massive enough (within one
or two decades of $M_{GUT}$) in order to leave
perturbative gauge coupling unification (marginally) intact.

Finally, even though $N = 11$ or $N = 18$ is plausibly large enough to
overcome the problem of negative third-generation scalar masses even in
high-scale theories, we note that the
new physics at the scale $M$ gives rise to third-generation scalar
masses of order $\de m_3 \sim \sqrt{cN} (\vu/M) M_{\rm comp}$, which
can be in the range $100\GeV$--$1\TeV$.
If this contribution is positive, it may improve the problem with the
negative log-enhanced contributions to the third-generation scalar
masses.
The contributions of new physics at the scale $M_*$ can give
the gravitino a mass of order $100\GeV$ in high-scale models,
so the gravitino need not be lightest \susc particle (LSP)
in these models.
The LSP is most likely a neutralino with a mass in the $100\GeV$ range,
which is a traditional favorite candidate for cold dark matter.

\subsection{Implications for Flavor Physics}
We now turn to the phenomenological implications of the models with
two composite generations, concentrating mainly on flavor physics.%
\footnote{Theories of flavor exploiting compositeness
(but not addressing \susy breaking) have been constructed in
\cite{KaplanSchmaltz}.}
If we assume that new physics at the scale $M$ is responsible for the
Yukawa couplings, then the Yukawa couplings will arise from operators
of the form \Eq{YukOp}.%
\footnote{Models with dynamical \susy breaking and composite states with
large global symmetries were also found in \Ref{LutyTerning}.
However, the composite states were high-dimension baryons,
and the Yukawa couplings for the composite generations are suppressed by
the ratio of the compositeness scale to the higher
scale $M$ raised to the  $30^{\rm th}$ power.
Therefore, these models cannot naturally produce large enough
Yukawa couplings even for the light generations.}
This gives rise to Yukawa matrix with the skeletal form
\beq
y \sim \pmatrix{\ep^2 & \ep^2 & \ep \cr
\ep^2 & \ep^2 & \ep \cr
\ep & \ep & 1 \cr},
\quad
\ep \sim \left( \frac{\vu}{M} \right)^2.
\eeq
It is clear that additional structure is needed to construct fully
realistic Yukawa matrices.
However, this is certainly a good starting point for constructing a
theory of flavor, and the automatic $\epsilon$ suppressions due
to the composite nature of the first two generations leave a milder
hierarchy in the coefficients of the higher-dimension operators
that needs to be explained.
For $\ep$ in the range $10^{-2}$ to $10^{-1}$, realistic fermion masses
can be obtained with simple textures and hierarchies of order $10$
in the effective coupling constants.

Let us turn to the issue of FCNC's due to non-degeneracy of the
scalar masses of the first two generations.
We emphasize again that, due to approximate flavor
symmetries of the strong dynamics, the leading contribution to the
soft masses is equal for the first two generations, and the issue
is whether sufficient degeneracy is maintained to
avoid FCNC constraints after the effects of the
higher dimension operators are included.%
\footnote{Flavor symmetries have been used to constrain both the form
of the Yukawa matrices and the scalar mass matrices, thereby addressing
both the \susc and usual flavor problems \cite{U(2),NS}.
In our case, however, the approximate flavor symmetry
guaranteeing scalar degeneracy need not be respected by the higher dimension
operators generating the Yukawa couplings.}
The size of the corrections depends on the flavor physics at the
scale $M$.
For example, we have already pointed out that it is possible that
the flavor physics has a GIM mechanism that suppresses FCNC's.
We now analyze the possibility that there is no alignment mechanism at
the scale $M$, so the off-diagonal scalar masses are suppressed only
by the powers of $\La / M$ that suppress the corresponding Yukawa
couplings.
The mixing contributions to the soft mass matrices come from
operators such as
\beq
\de\scr{L} \sim \myint d^4\th \left[
\frac{c}{M^2} (A \bar{U})^\dagger (A \bar{U}) \right],
\eeq
which give
\beq\eql{compsupp}
\frac{\de m_{jk}^2}{M_{\rm comp}^2}
\sim c \left( \frac{\vu}{M} \right)^2
\sim c \sqrt{y_{jk}}.
\eeq
(Note that the operator of \Eq{softmassbreak} is enhanced by a
factor of $N$, but is flavor-diagonal.)
The most stringent FCNC bounds come from the $K$--$\bar{K}$ system,
and can be summarized as
\beq
\mbox{Re}\left(\frac{\de m_{\tilde{d}\tilde{s}}^2}{M_{\rm comp}^2}
\right) \lsim 10^{-1} \frac{M_{\rm comp}}{10\TeV}, \quad
\mbox{Im} \left(\frac{\de m_{\tilde{d}\tilde{s}}^2}{M_{\rm comp}^2}
\right) \lsim 10^{-2} \frac{M_{\rm comp}}{10\TeV}.
\eeq
The constraint from Re($\de m^2_{\tilde{d}\tilde{s}}$) gives
(using $y_{ds} \sim \sqrt{y_d y_s}$)
\beq\eql{cbound}
c \lsim 5\, \frac{M_{\rm comp}}{10\TeV},
\eeq
which is plausibly satisfied for $M_{\rm comp}$ as low as $1\TeV$
given the uncertainties.
In order to also evade the bounds from
$\Im(\de m^2_{\tilde{d}\tilde{s}})$, we must assume that the
\CP-violating phase in this quantity is somewhat small, of order
$\frac{1}{10}$.
Alternately, $M_{\rm comp} \sim 10\TeV$ is completely safe from
all constraints.

Even if the induced non-degeneracies between the first two
generation sfermions are small enough to avoid present FCNC constraints,
there is still a rich spectrum of flavor changing signals due
to the non-degeneracy between the first two and third generation
sfermions.
If the sfermion mixing angles are CKM-like, flavor-violating
signals are expected at experimentally interesting levels in a wide
variety of processes such as $\mu \to e \gamma$, $\mu \to 3 e$,
$B$--$\bar{B}$ mixing, and electron/neutron electric dipole moments
\cite{BarbHall}.

Finally, we note that the new physics at the scale $M$ may provide a
solution \cite{GMmu} to the ``$\mu$ problem.''
If the low-energy theory contains the terms
\beq\eql{muterm}
\de\scr{L} = \myint d^4\th\, \frac{c'}{M} \tr(\bar{U}^\dagger \bar{U})
( H \bar{H} + \hc),
\eeq
where $H$, $\bar{H}$ are the standard-model Higgs fields, then the
low-energy theory contains $\mu$ and $B\mu$ terms of order
\beq
\mu \sim c' N \frac{\avg{F_{\bar{U}}} \vu}{M^2}
\sim c' N \sqrt{y} M_{\rm comp},
\quad
B\mu \sim c' N \frac{\avg{F_{\bar{U}}}^2}{M^2}
\sim c' N \sqrt{y} M_{\rm comp}^2,
\eeq
where $y \sim (\vu / M)^4$ is the magnitude of a Yukawa coupling
generated at the scale $M$.
If we want $\mu^2 \sim B\mu$, then we need $c' N \sim 1/\sqrt{y}$,
which is plausible for large $N$.
(The parameter $c'$ is of order 1 or $1/N$, as in the discussion
below \Eq{seec}.)
In this case, both $\mu$ and $B\mu$ are naturally near $M_{\rm comp}$,
which is somewhat large for electroweak symmetry breaking even if
$M_{\rm comp} \sim 1\TeV$.
However, given the large uncertainties and model-dependence in these
estimates, this mechanism may work in a detailed model.

\subsection{Decay of the False Vacuum}
All of the models above require that the universe live in a false
vacuum on the ``baryon'' branch, and so we must consider the
possibility of the decay of the vacuum.
All of the \susc vacua occur at infinite field values on other
branches of the moduli space.
Therefore, the energy difference between the false vacuum and the
true vacuum is small compared to the distance in field space to the
classical escape point.
We can therefore give a conservative bound on the tunneling rate
by approximating the potential as completely flat.
In that case, the Euclidean tunneling action is
\cite{flatbounce}
\beq
S_{\rm tunnel} \simeq 2\pi^2 \frac{(\De\phi)^4}{V},
\eeq
where $\De\phi$ is the distance in field space to the classical escape
point, and $V \sim \avg{F_{\bar{U}}}^2$ is the value of the energy density
in the false vacuum.
Since $(\De\phi)^2 \gg \avg{F_{\bar{U}}}$ in our models,
this always gives a negligible tunneling rate.

\section{Discussion and Conclusions}
We have presented new models of dynamical \susy
breaking in which the same strong dynamics breaks \susy
and gives rise to massless composite fermions that we
identify with quarks and leptons of the first two generations.
Since the corresponding composite squarks and sleptons arise
directly from the \susy breaking sector, they receive \susy-breaking
soft masses directly, without ``mediation'' via gravitational or
SM gauge interactions.
In this sense, these models provide an alternative to the
``modular'' structure of realistic models of \susy breaking,
where \susy is broken in a separate sector of the model and
communicated by messenger interactions to the observed particles.

It is also pleasing that the models we construct are quite simple.
As an illustration, we write the complete $N = 18$ model below.
The gauge group is
\beq
\SU{4} \times \SU{18} \times \left[ \SU{18} \right]
\eeq
where $\SU{5}_{\rm SM}$ (the usual embedding of the standard-model group)
is embedded into $[\SU{18}]$ so that ${\bf 18}
\to {\bf \bar{5}} + {\bf 10} + (3 \times {\bf 1})$.
The field content is
\beq\bal
Q &\sim (\fund, \fund, {\bf 1})~,
\\
L &\sim (\bar{\fund}, {\bf 1}, \fund)~,
\\
\bar{U} &\sim ({\bf 1}, \bar{\fund}, \bar{\fund})~,
\\
A &\sim ({\bf 1}, \asymm, {\bf 1})~,
\eal\eeq
together with a single (third) generation
$\Phi_{\mybar{\bf 5}}$, $\Phi_{\bf 10}$ and
Higgs fields $H$, $\bar{H}$.
The model has a superpotential of the form
\beq\bal
W &\sim L Q \bar{U} + H \Phi \Phi
+ \frac{1}{M^2} (A\bar{U}^2) H \Phi
\\
&\quad
+ \frac{1}{M^3} (A\bar{U}^2) (A\bar{U}^2)
+ \frac{1}{M^4} (A\bar{U}^2) (A\bar{U}^2) H
\eal\eeq
where we have omitted indices for simplicity.
The higher-dimension operators generate Yukawa couplings involving
the composite states and eliminate unwanted composite fermions
from the low-energy spectrum.
This model generates two composite
generations of quarks and leptons with small Yukawa couplings,
breaks \susy, communicates \susy breaking directly to the
composite squarks and sleptons, and gives sufficiently large
gaugino masses through gauge loops.

It is striking that a simple model such as this can be completely
realistic, with the compositeness scale ranging anywhere from
$10\TeV$ to the Planck scale.
The leading contribution to the scalar masses is naturally flavor-diagonal
due to an approximate symmetry of the strong dynamics that is present
even if $\la$ has arbitrary flavor structure;
this symmetry is violated only by ``perturbative'' corrections of order
$\la^2 / (16\pi^2) \sim 10^{-4}$.
These global symmetries also lead to the striking prediction that
(depending on the model) some or all of the scalar masses of the first
two generations  unify at the compositeness scale, which need not be close
to the GUT scale.
(Models with flavor symmetries can also predict scalar
unification at some level, but they cannot naturally explain unification
between scalars with different gauge quantum numbers below the
GUT scale.%
\footnote{Even if the scalar unification scale is close to the GUT scale,
the model above predicts that \emph{all} squarks and sleptons from the first
two generations unify.
Even the degeneracy of squark and slepton masses within a generation
is not easy to understand in $\SO{10}$, since it is broken by $D$
terms corresponding to broken generators.})
We emphasize that these features are present in our model without the
need to impose any flavor symmetry on the underlying theory.

The Yukawa couplings are generated by new physics at a scale
above the compositeness scale, naturally explaining why the fermion
masses of the first two generations are small, while the corresponding
scalar masses are large.
In the absence of any flavor alignment mechanism,
the off-diagonal terms are just
compatible with existing constraints on \CP-conserving FCNC's
if the scalar masses are in the $1\TeV$ range.
(Consistency with $\ep_K$ requires scalar masses of order $10\TeV$.)
In either case, one expects FCNC's that may be
observed with increased experimental sensitivity.
The models require a dynamical assumption regarding the sign of a
dynamically-generated mass term.
(If the sign is opposite to what is assumed here, one can use the
dynamics to build a composite messenger model of direct gauge-mediated
\susy breaking along the lines of
\Ref{LutyTerning}.)

We close with some speculations on how to build more attractive models
based on the ideas presented here.
The models discussed in this paper have a large number of states
charged under the standard-model gauge group above the compositeness
scale, resulting in a Landau pole close to the compositeness scale.
Also, the scale of flavor physics must be very close to the compositeness
scale in order to generate sufficiently large Yukawa couplings.
Both of these potential difficulties may be alleviated if one could
find models where the composite states correspond to dimension-2
``meson'' operators of the form $P_1 P_2$, where $P_{1,2}$ are
strongly-coupled preons.
In that case, Yukawa couplings involving the composite states arise
from terms in the Lagrangian of the form
\beq
\de\scr{L} \sim \myint d^2\th \left[ \frac{1}{M^2} (P_1 P_2)^2 H
+ \frac{1}{M} (P_1 P_2) \Phi H \right] + \hc,
\eeq
where $H$ is an elementary Higgs field and $\Phi$ is an elementary
third-generation quark or lepton field.
This gives rise to Yukawa couplings for the composite fields of
order
\beq
y_q \sim \frac{\avg{P}^2}{M^2}
\eeq
compared to $\avg{P}^4 / M^4$ in our models.
This would allow the flavor scale to be larger compared to the
compositeness scale.
Off-diagonal scalar mass terms for the composite fields arise from
\beq
\de\scr{L} \sim \myint d^4\th\, \frac{1}{M^2}
(P_1^\dagger P^{\vphantom{\dagger}}_1)
(P_2^\dagger P^{\vphantom{\dagger}}_2),
\eeq
giving rise to off-diagonal scalar masses for the composite states
\beq
\frac{\de m^2_{jk}}{M_{\rm comp}^2}
\sim \frac{\avg{F_{P}}^2}{M^2} \sim \frac{\avg{P}^2}{M^2} M_{\rm comp}^2
\sim y_{jk} M_{\rm comp}^2,
\eeq
where $y_{jk}$ is the corresponding off-diagonal Yukawa coupling.
This is an extra suppression by $\sqrt{y_{jk}}$ compared to our
models, which makes FCNC's completely safe.
Finally, one might hope that the group-theory structure of such models
allows more economical models with a higher Landau pole for the
standard-model interactions.
It is also interesting to see if models of this type can give rise
to realistic theories of flavor.
We believe that these are promising directions, and work along these
lines is in progress \cite{wip}.

\section{Acknowledgments}
N.A-H. thanks L. Hall and H. Murayama for useful discussions.
M.A.L. thanks the theory group at LBNL for hospitality
during the initial stages of this work.
J.T. thanks M. Schmaltz for useful discussions.
N.A-H. is supported by the Department of Energy under contract
DE-AC03-76SF00515.
M.A.L. is supported by a fellowship from the Alfred P. Sloan Foundation.
J.T. is supported by the National Science
Foundation under grant PHY-95-14797, and is also partially supported by
the Department of Energy under contract DE-AC03-76SF00098.



\end{document}